\documentclass[twocolumn,showpacs,preprintnumbers,amsmath,amssymb,prb,aps]{revtex4-1}
\usepackage{epsfig}
\usepackage{epstopdf}
\usepackage{graphicx}
\usepackage{dcolumn}
\usepackage{bm}
\usepackage{slashbox}
\usepackage{textcomp}
\usepackage{array}
\usepackage{subcaption}
\usepackage{booktabs}

\begin{document}

\title{Effect of density functionals on the vibrational and thermodynamic properties of Fe$_{2}$VAl and Fe$_{2}$TiSn compounds}
\author{Shivprasad S. Shastri}
\altaffiliation{Electronic mail: shastri1992@gmail.com}
\author{Sudhir K. Pandey}
\affiliation{School of Engineering, Indian Institute of
Technology Mandi, Kamand - 175005, India}

\date{\today}

\begin{abstract}
First-principles phonon calculations along with Kohn-Sham density functional theory (DFT) is an essential tool to study the lattice dynamics, thermodynamical properties and phase-transitions of materials. The two full-Heusler compounds Fe$_{2}$VAl and Fe$_{2}$TiSn are studied for lattice vibration dependent properties using finite displacement method and supercell approach. For the investigation, four density functionals viz., LDA, PBE, PBEsol and meta-GGA SCAN are employed. Using these functionals, phonon dispersion, phonon density of states (DOS), partial density of states (PDOS) thermal propertis and zero-point energy are calculated at equilibrum lattice parameters under harmonic approximation. For the two compounds the Debye temperatures are calculated from the obtained phonon DOS which are $\sim$660 K and $\sim$540 K, respectively. The obtained results from different functionals are compared among each other. The overall phonon energy in the dispersion is found to be $\sim$15 meV higher in Fe$_{2}$VAl than the Fe$_{2}$TiSn compounds. For the two compounds PBE is yielding the lowest phonon frequencies while LDA or SCAN functional is giving the highest. The same pattern is observed in phonon DOS plots of two compounds. The zero-point energy calculated is the highest from SCAN (21.04 and 16.95 J) and the lowest from PBE functionals (20.09 and 16.02 J) obeying the same trend as frequency for both compounds. A general prediction of nature of lattice thermal conductivity is made based on the velocities of acoustic phonons which is in agreement with the qualitative behavior of reported experimental thermal conductivity of two compounds. Phonon spectra obtained from PBE and SCAN have similar general features while those from LDA and PBEsol have resembling features for Fe$_{2}$VAl, while this trend is not observed for the compound Fe$_{2}$TiSn. 

\end{abstract}

\maketitle

\section{Introduction} 
First-principles Kohn-Sham density functional theory (DFT)\cite{kohnsham} is one of the most widely used theoretical investigation tools in materials science. This tool has been very helpful since it's inception for the study of electronic structure, and other ground-state properties. This is also made possible with the aid of advancements in the field of high-perfomance computing and it's cost effectiveness.\cite{skelton2015} In the KS-DFT the total ground-state enrergy is calculated as a functional of the ground-state electron density for a static lattice and is minimized with respect to lattice constant. The total ground state energy is the sum of kinetic energy, Coulomb energy and exchange-correlation (XC) energy. And this XC energy is the part which is approximated in practice.\cite{hao2012lattice} So, the best explanation to experimental results from DFT calculations depends on the approximation used in XC functionals. There are large number of approximations available to XC energy. LDA and GGA-PBE are the mostly used density functionals. 

This application of first-principles calculations is extended in the field of condensed matter physics and materials science by the introduction of phonon calclations.\cite{togo,wangnature,skelton2015} When an atom is displaced from the equilibrium position in a crystal forces arise on all the atoms. By calculating the force on the each atom using the total energy of the crystal, phonon frequencies can be obtained. This approach of calculation is called finite-displacement method (FDM). Another method uses density functional perturbation theory (DFPT) to calculate the phonon frequencies and hence dependent properties.\cite{togo}

Fe$_{2}$VAl and Fe$_{2}$TiSn are the two compounds belonging to the family of full-Heusler alloys. These compounds are reported to have non-magnetic ground state with applications in many fields. They crystallise in cubic $L2_{1}$ phase with space group $Fm-3m$.\cite{nishino06,yabucchi,galanakis} The two compounds are also being studied for the application as thermoeletrics for heat energy conversion.\cite{luethermalfe2tisn,luefe2valthermoelectric} A thermoelectric is characterized by it's figure of metit (ZT) value. As it is well known that ZT is proportional to the electrical conductivity and inversly related to thermal conductivity, it is important to study the electronic structure of a material in order to understand or improve the efficiency.\cite{mahan1989figure,mahan2016introduction} Since, the thermal conductivity has an important contribution from  lattice vibrations, it is must to investigate the lattice thermal conductivity if one has to properly understand a thermoelectric in order to tailor it's efficiency.\cite{mahan2016introduction}

Keeping this in mind, we have studied the electronic strucure of the two compounds employing five XC functionals in our recently reported study.\cite{Shastri2018} We also calculated effective masses for two compounds and found that Fe$_{2}$TiSn showing large value of effective mass having flat band. In this direction, to completely understand the thermoelectric behavior, there is a need to study the lattice part of thermal conductivity of the two compounds through phonon calculations. Also, the lattice dynamics study of the two compounds in literature using different functionals is missing. Performing the test of functionals to study phonon frequencies along with electronic structure gives strict test of the functional.\cite{skelton2015}

Therefore, in the present work, we have undertaken the study of two full-Heusler compounds, Fe$_{2}$VAl and Fe$_{2}$TiSn using four XC  functionals. The XC functionals employed are, LDA-PW92\cite{lda92}, GGA-PBE\cite{pbe}, PBEsol\cite{pbesol} and SCAN\cite{scan} meta-GGA. Using these functionals phonon dispersion, density of states (DOS), partial density of sates (PDOS) are calculated at equilibrium lattice parameters under harmonic approximation. Further, variation in thermodynamical quantities, constant volume specific heat $C_{V}$ and Helmholtz free energy (F) as a function of temperature are calculated. The Debye temeperature and the zero-point energy contribution to the static lattice are also calculated. The effect of the various functionals on the obtained phonon properties are discussed. Lattice thermal conductivity prediction from the slope of acoustic phonon branches are guessed. And this result is compared qualitatively with the experimental thermal conductivity study of two compounds.
  
\section{Computational details}
In this work calculations are carried out using two computational programs. To obtain the total forces on the atoms , first-principles full-potential linearized augmented plane wave (FP-LAPW) based program WIEN2k\cite{wien2k} is used as force calculator. Self-consistent cycles are run till the sum of total forces on atoms is less than 0.1 mRy/bohr. The calculations are performed using four exchange-correlation functionals viz., LDA\cite{lda92}, PBE\cite{pbe}, PBEsol\cite{pbesol} and recent meta-GGA functional SCAN.\cite{scan} A k-mesh of size 5x5x5 is used for both Fe$_{2}$VAl and Fe$_{2}$TiSn compounds for force calculation. For the phonon calculations part PHONOPY\cite{togo} code is used. PHONOPY\cite{togo} is a first-principles phonon calculations tool which can handle force constants obtained from both finite displacement method (FDM) and density functional pertrubation theory (DFPT) for phonon frequency analysis. Here, FDM and supercell approach is used to calculate phonon properties. A supercell of size 2x2x2 of primitive lattice type is constructed for both the compounds with 128 atoms. The supercell is used to obtain the long-range force constants between atoms.\cite{skelton2014thermal}
 An artificial force is created in the systems by displacing atom of each kind (Fe, V, Al, Ti, Sn) in the formula units by a distance of 0.02 bohr in x-direction. Further analysis of total forces on atoms to get force-cosntants and post-processing of them are carried out using PHONOPY\cite{togo} to obtain phonon dispersions and phonon density of states (DOS) and thermal properties on a sampling mesh of 21x21x21. 
 Fe$_{2}$VAl and Fe$_{2}$TiSn full-Heusler compounds have cubic \textit{L2$_{1}$} crystal structure with Fm-3m space group. The Fe atoms occupy the Wyckoff position 8c $(\frac{1}{4},\frac{1}{4},\frac{1}{4})$, Y atoms occupy Wyckoff  position 4a $(0,0,0)$ and Z atoms occupy Wyckoff position 4b $(\frac{1}{2},\frac{1}{2},\frac{1}{2})$.\cite{mahanti} Where Y=V,Ti and Z=Al,Sn in the general representation Fe$_{2}$YZ of these two full-Heusler compounds. \cite{Shastri2018}

\section{Results and Discussion}
The lattice vibration dependent properties of the two Heusler compounds are studied under harmonic approximation at equilibrium lattice parameters. The equilibrium lattice constants are obtained by minimizing the total energy with volume. The obtained optimized lattice constant values from four different XC functionals are presented in Table 1. The details of the volume-optimization calculation can be found in our previous work.\cite{Shastri2018}

\begin{table*}
\caption{Calculated lattice constants $a_{0}$ for the two compounds using five exchange-correlational functionals}
\resizebox{0.6\textwidth}{!}{%
\begin{tabular}{@{\extracolsep{\fill}}c c c c} 
 \hline\hline
 & \multicolumn{1}{c}{\textbf{Fe$_{2}$VAl}} & & \multicolumn{1}{c}{\textbf{Fe$_{2}$TiSn}}\\
 \cline{2-2} \cline{4-4}
       & Lattice constant ($a_{0}$) & & Lattice constant ($a_{0}$)\\
       & ({\AA}) & & ({\AA})\\
 \hline
LDA   & 5.5955 & &  5.9102 \\
PBE   & 5.7089  &  &6.0423 \\
PBEsol&  5.6478  & &5.9664 \\
SCAN&  5.6509  &  &5.9762 \\
 \hline\hline
\end{tabular}}
\end{table*} 

From the Table 1.\cite{Shastri2018} the trends in the obtained equilibrum lattice parameters from four functionals can be observed. In these cases also the overbinding nature of LDA and relatively underbinding nature of PBE are clearly exhibited with the lowest and highest values of lattice constants, respectively.

To further study the lattice stability, phonon energy and thermodynamic properties of the two full Heusler compounds, first-principles phonon calculations are carried out. 

\subsection{\label{sec:level2}Vibrational properties}
The phonon band structure and density of states (DOS) and partial density of states (PDOS) are calculated for the two compounds using four XC functionals. The corresponding phonon band structures from the four functionals for Fe$_{2}$VAl and Fe$_{2}$TiSn are shown in Fig. 1(a)-(d) and (e)-(h), respectively. The vibrational spectra is plotted along $\Gamma$-$X$-$W$-$\Gamma$-$L$ directions in the first Brillouin zone. A crystal is considered to be mechanically stable if the potential energy of the crystal increases for any displacements of atoms inside it. According to harmonic approximation, this condition is satisfied when all the phonons have real and positive frequencies (or energies).\cite{togo} The negative or imaginary frequencies imply that the crystal is dynamically unstable discarding the possibility of synthesis in that phase. In our case, the obtained dispersion curves for two compounds does not show any negative frequecies indicating that both the compounds are stable in the cubic \textit{L2$_{1}$} phase. The two branches out of the total three acoustic branches are doubly degenerate (or nearly doubly degenerate) nearly half-way along $\Gamma$-$W$ direction and fully along $\Gamma$-$L$ direction in case of both the compounds. The acoustic branches of Fe$_{2}$VAl are following the linear relation with \textbf{q} for more distance (i.e more shorter wavelength) than in the case Fe$_{2}$TiSn, along both $\Gamma$-$W$ and $\Gamma$-$L$ directions. In this linear region the group velocity will be the same as phase velocity.\cite{ashcroft} Few of the acoustic and optical branches of Fe$_{2}$VAl are degenerate at an energy of $\sim$29 meV and $\sim$33 meV, respectively at the zone boundary containg point $W$. But this nature is not observed in case of Fe$_{2}$TiSn vibrational spectrum, where there is a clear seperation between optical and acoustic phonon branches. On observing the energy scales of the dispersion curves of the two compounds, it is visible that the vibrational energy of phonons is more in Fe$_{2}$VAl than in the case of Fe$_{2}$TiSn compound. The Fe$_{2}$TiSn compound contains heavier element Sn and it's atomic mass is more than the Fe$_{2}$VAl compound. As the vibrational frequency, in turn energy, is inversly proportional to the suare root of mass of elements in the compound, these observations justify the frequency-mass relation. In the Fe$_{2}$VAl vibrational spectrum we observe three acoustic branches are crossing with some of the optical branches. And there are three high energy optical branches of $\sim$50 meV with a distinct seperation from the other branches. In Fe$_{2}$VAl, looking at the atomic masses of the constituents of the formula unit, Al atom has the lowest atomic mass. Therefore, the major contribution to these curves should be from that atom.\cite{kanchana} The crossing branches in the region between $\sim$25-40 meV energy are mainly due to the near atomic mass elements Fe and V. Now, while going from Fe$_{2}$VAl to Fe$_{2}$TiSn the following changes in the vibrational spectrum of the latter compound are observed. The highest energy optical phonon branches are observed $\sim$35 meV in the energy scale and which is shifted down compared to that of Fe$_{2}$VAl. Compared with the former compound here, three lower energy acoustic branches are found to be well seperated from the higher energy optical phonon disperion curves except at the $L$-point with energy difference of $\sim$1 meV. In this case, the major contribution to the energy of these acoustic phonons are from the heavier atomic mass element Sn in the compound. The contribution to the optical branches in $sim$25-40 meV regions are maninly due to the near atomic mass elements Fe and Ti atoms in Fe$_{2}$TiSn. These qualitative explanations are justified further through phonon PDOS plots in figure 3. As mentioned before, the linear relationship bewteen frequency (or energy) and \textbf{q} of acoustic branches is followed for small values of \textbf{q} in Fe$_{2}$TiSn compared to that in the dispersion relation of Fe$_{2}$VAl. With the seperation between acoustic and optical phonons mentioned, the degeneracy observed among them in case of the former compound is absent here. One of the interesting features of the vibrational spectra of the two compounds can be highlighted here. That is, on examining the slopes of the acoustic branches of the two compounds, the slope values in the linear region are decreasing from Fe$_{2}$VAl to Fe$_{2}$TiSn. It is known that, slopes of the acoustic branches are associated with sound velocity and in turn, give the group velocity (and phase velocity in linear region). Also, lattice thermal conductivity in a solid is directly proportional to the sound velocity.\cite{maciathermoelectric,ashcroft} So, knowing the value of sound velocity one can proceed to calculate lattice thermal conductivity. Qualitatively, thus it can be said that lattice thermal conductivity is higher in Fe$_{2}$VAl compared to Fe$_{2}$TiSn which is in the context of thermoelectric application is a useful information. Lue et. al have reported the thermoelectric properties of Heusler compounds Fe$_{2-x}$V$_{1+x}$M (M=Al,Ga)\cite{luefe2valthermoelectric} and Fe$_{2-x}$Ti$_{1+x}$Sn.\cite{luethermalfe2tisn} The reported value of thermal conductivity is higher in case of Fe$_{2}$VAl than in the Fe$_{2}$TiSn compound under the temperature range studied. Also, in both the compounds they have suggested total thermal conductivity is essentially from lattice part of thermal conductivity. These experimental results are in support of the qualitative explanation given above from our study of the two compounds with sound velocity higher in the former compound than in the Fe$_{2}$TiSn as stated before.
 Thus, generally, based on the phonon disprersion analysis of this class of compounds we can say that going for a higher atomic mass compounds in the series can leads to lower value of lattice thermal conductivity keeping in mind the corresponding electronic structure of the compound with the supporting data of experimental thermal conductivity from Fe$_{2}$VAl, Fe$_{2}$VGa\cite{luefe2valthermoelectric}, Fe$_{2}$TiSn compounds.\cite{luethermalfe2tisn}
We could not come across any neutron inelastic scattering experimental data of phonon dispersion of both the compounds to compare the vibrational sperctrum of our study. Thus, it would be interesting to study the two compounds experimentally to compare the phonon dispersion results obtained from the calculations.

\begin{figure*}
 
\begin{subfigure}{0.48\textwidth}
\includegraphics[width=0.9\linewidth, height=6.0cm]{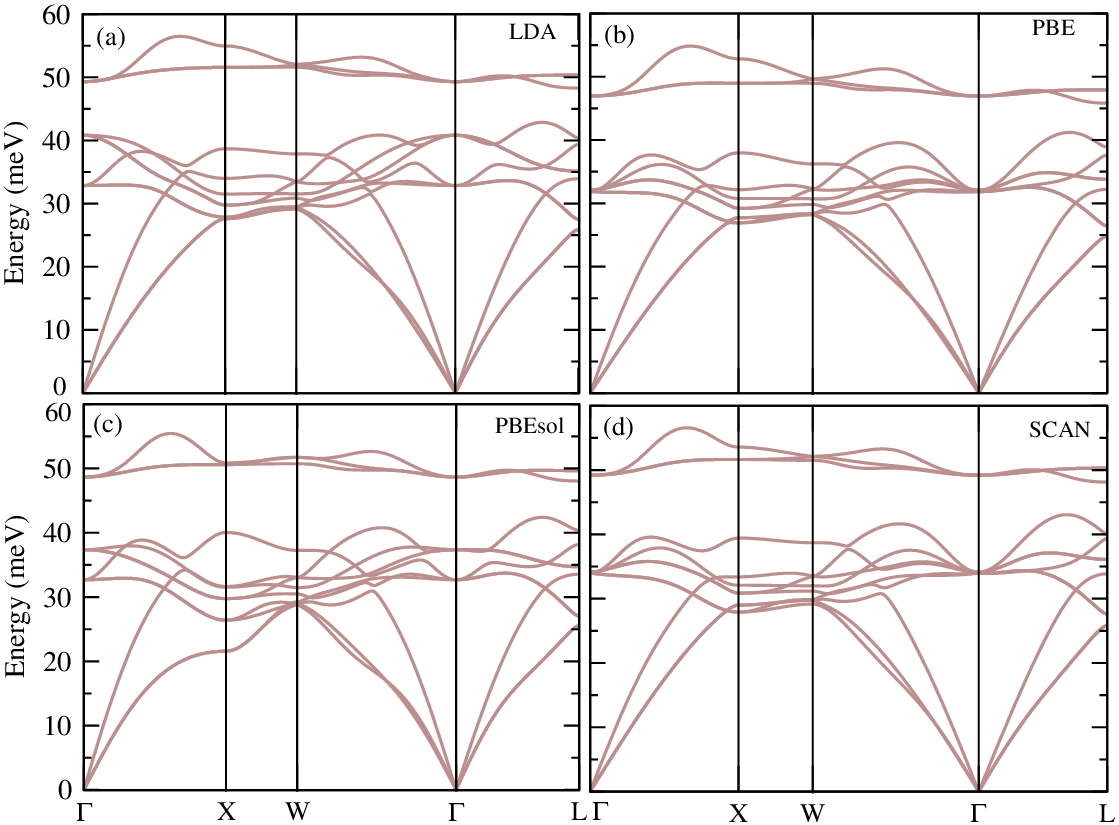} 
\end{subfigure}
\begin{subfigure}{0.48\textwidth}
\includegraphics[width=0.9\linewidth, height=6.0cm]{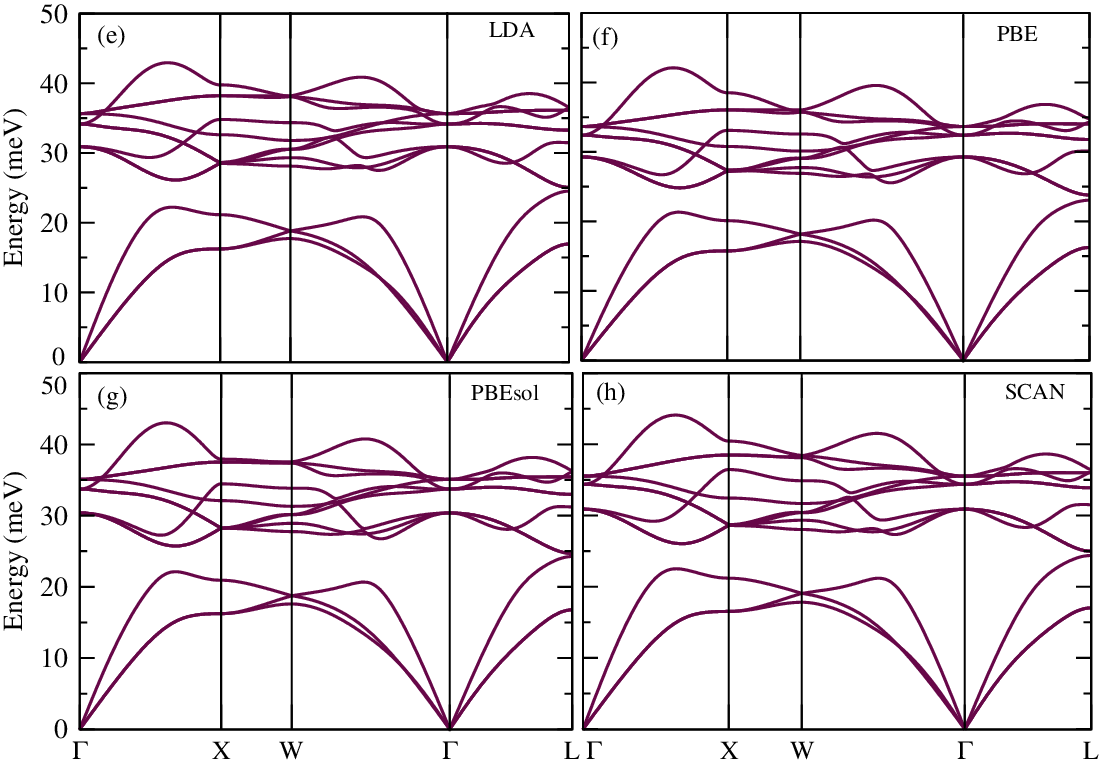}
\end{subfigure} 
\caption{Phonon dispersion curves for (a)-(d) Fe$_{2}$VAl and (e)-(h) Fe$_{2}$TiSn compounds}
\label{fig:image2}
\end{figure*}

\begin{figure*}
 
\begin{subfigure}{0.48\textwidth}
\includegraphics[width=0.9\linewidth, height=6.0cm]{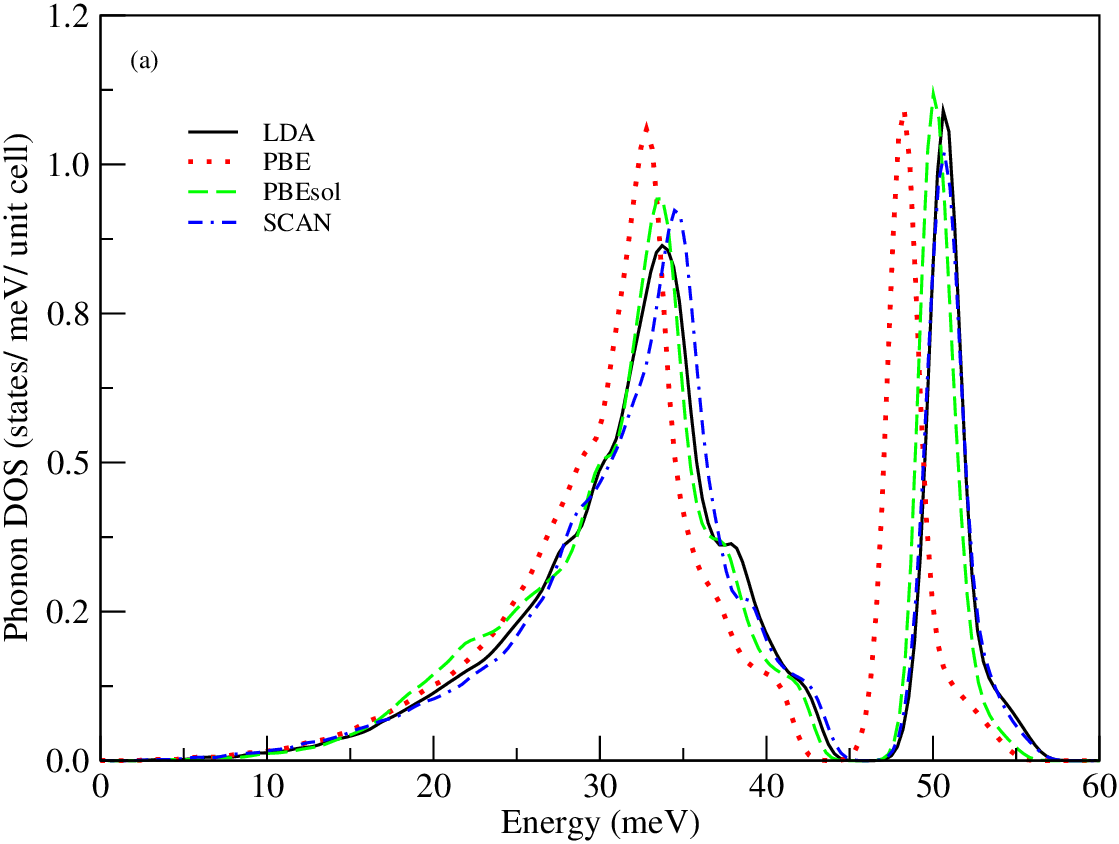} 
\end{subfigure}
\begin{subfigure}{0.48\textwidth}
\includegraphics[width=0.9\linewidth, height=6.0cm]{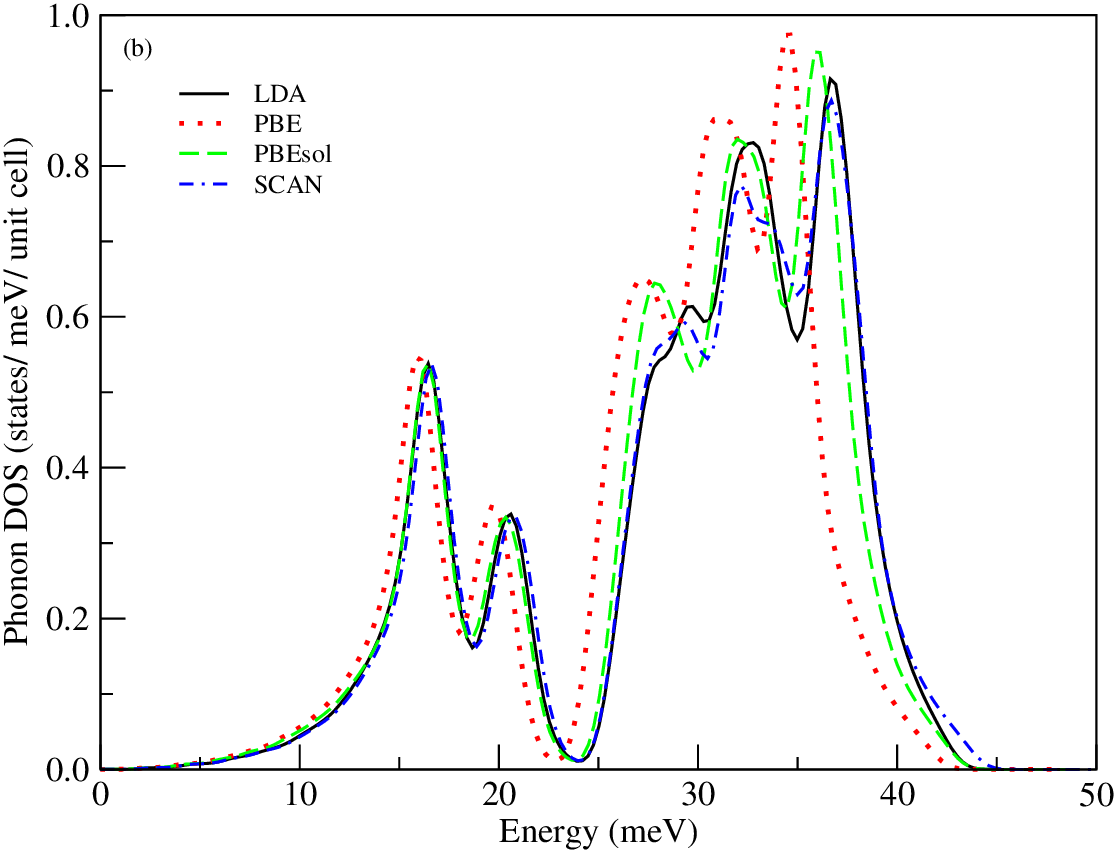}
\end{subfigure} 
\caption{Phonon DOS for (a) Fe$_{2}$VAl and (b) Fe$_{2}$TiSn compounds}
\label{fig:image2}
\end{figure*}

The phonon density of states (DOS) plots for two compounds are presented in Fig. 2 (a) and (b) from all four functionals under study. In Fig. 2(a), phonon DOS of Fe$_{2}$VAl can be seen. Corresponding to the crossing branches of acoustic and optical phonons of the compound high value of phonon DOS peak can be seen in $\sim$25-40 meV region. The highest energy DOS peak in the neighborhood of 50 meV is representing the seperated optical branches in the spectra which is contributed mainly from Al vibrational states.\cite{kanchana} Vibrational DOS of Fe$_{2}$TiSn compound is shown in Fig. 2(b). Clearly, the two peaks in the lower energy region correspond to the three acoustic branches in the disperion plot. The next seperated higher energy peaks represent the coupled optical phonon branches. The nature of the respective phonon DOS peaks justifies the phonon dispersion of the two compounds shown in Fig. 1.

To see the contribution to the phonon frequencies from the different atoms, phonon partial density of states are calculated. Fig. 3 (a)-(d) shows the partial DOS for Fe$_{2}$VAl obtained from four functionals. It is known that, Al is the lighter element compared to Fe and V. In the $\sim$20-40 meV region the main contribution to DOS is from Fe and V atoms. It can be observed that the higher energy three optical branches has primary contribution from Al atom. For Fe$_{2}$TiSn compound, the phonon partial DOS plot is given in Fig. 3 (e)-(h). In the plot the acoustic phonons belong to the energy upto $\sim$25 meV. As we can see, energies of acoustic phonon branches are cheifly from the heavier Sn atoms. Contribution to DOS from Fe and Ti atoms are less in this region. The PDOS plot also shows for Fe$_{2}$TiSn ,the optical phonons are contributed from the lighter Fe and Ti atoms. In both the compounds, the lighter atoms Al and Ti are contributing less in the lower frequency region and more in higher frequency region to the DOS, respectively. 

\begin{figure*}
 
\begin{subfigure}{0.48\textwidth}
\includegraphics[width=0.9\linewidth, height=6.0cm]{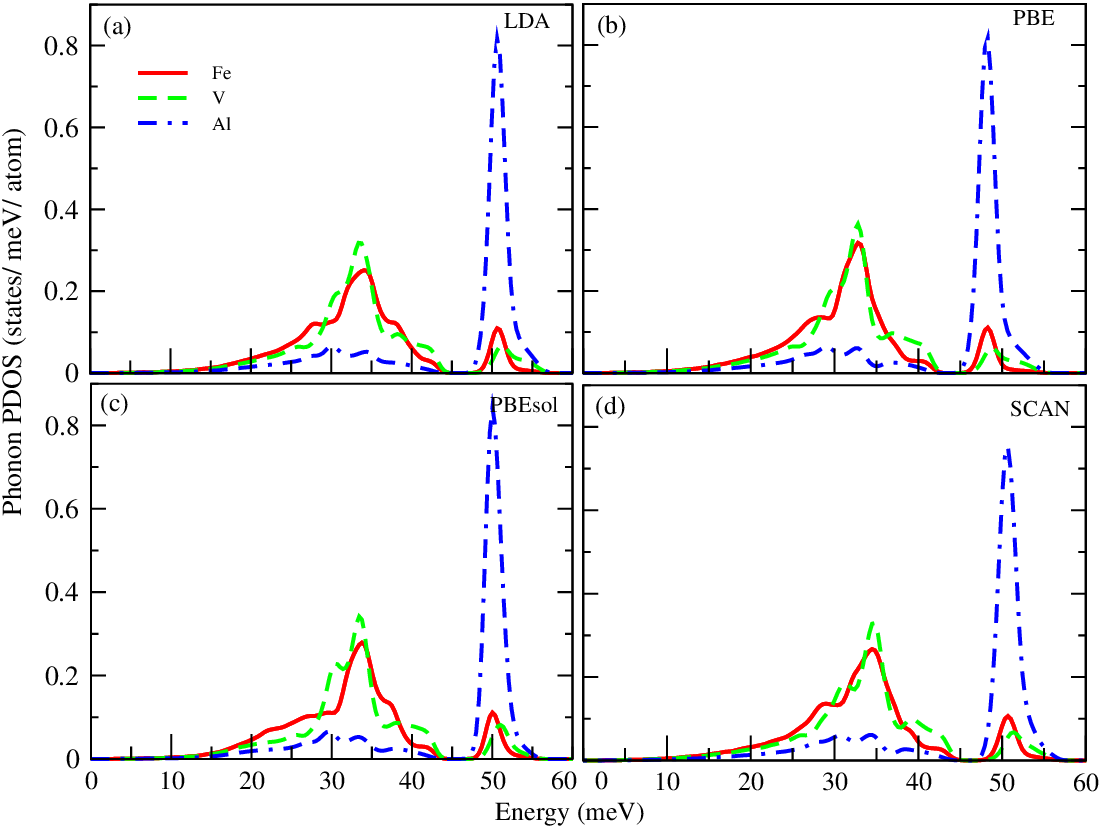} 
\end{subfigure}
\begin{subfigure}{0.48\textwidth}
\includegraphics[width=0.9\linewidth, height=6.0cm]{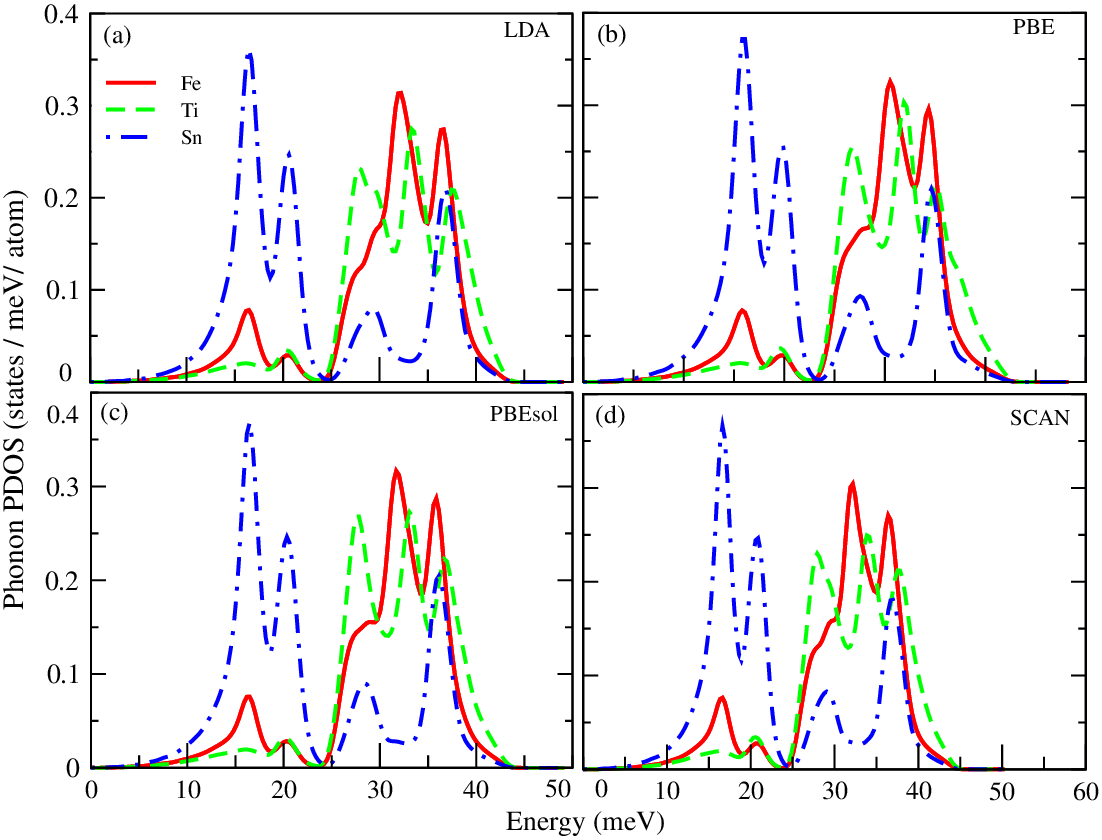}
\end{subfigure} 
\caption{Phonon partial density of states (PDOS) plots (a)-(d) Fe$_{2}$VAl and (e)-(h) Fe$_{2}$TiSn compounds}
\label{fig:image2}
\end{figure*}

\subsection{\label{sec:level2}Thermal properties}

\begin{figure*}
 
\begin{subfigure}{0.48\textwidth}
\includegraphics[width=0.9\linewidth, height=6.0cm]{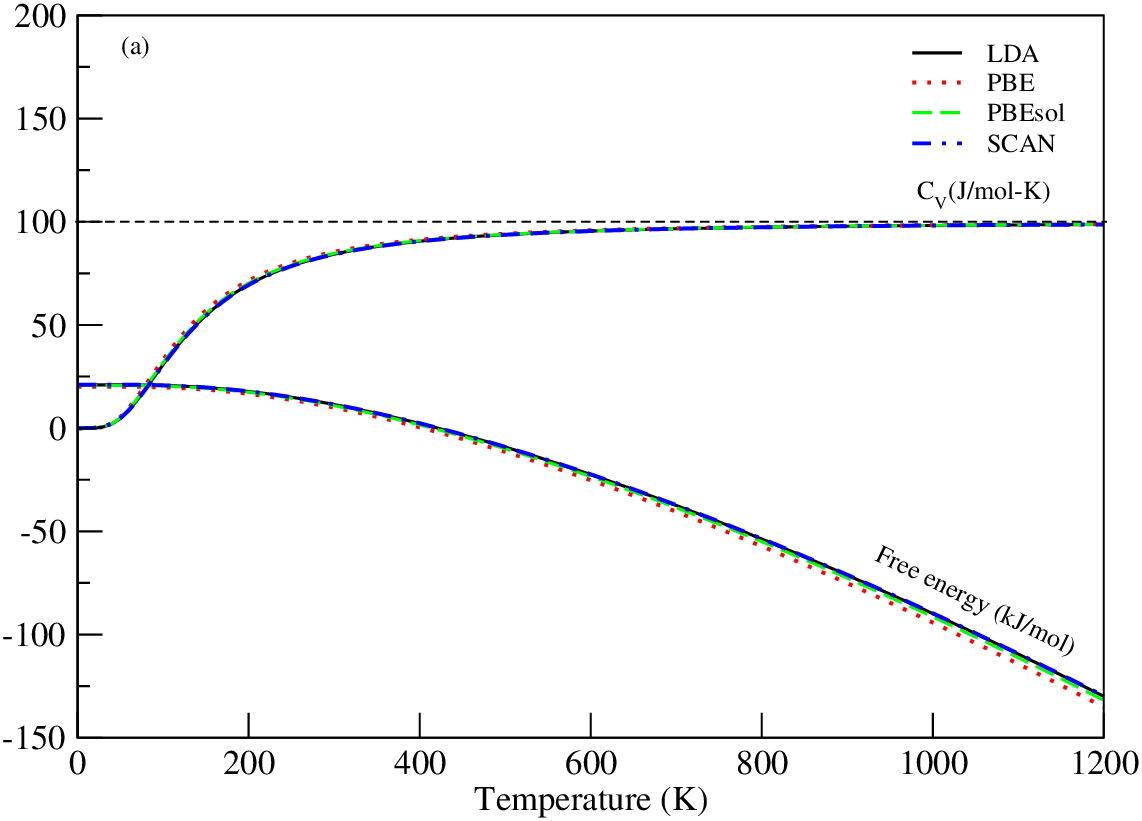} 
\end{subfigure}
\begin{subfigure}{0.48\textwidth}
\includegraphics[width=0.9\linewidth, height=6.0cm]{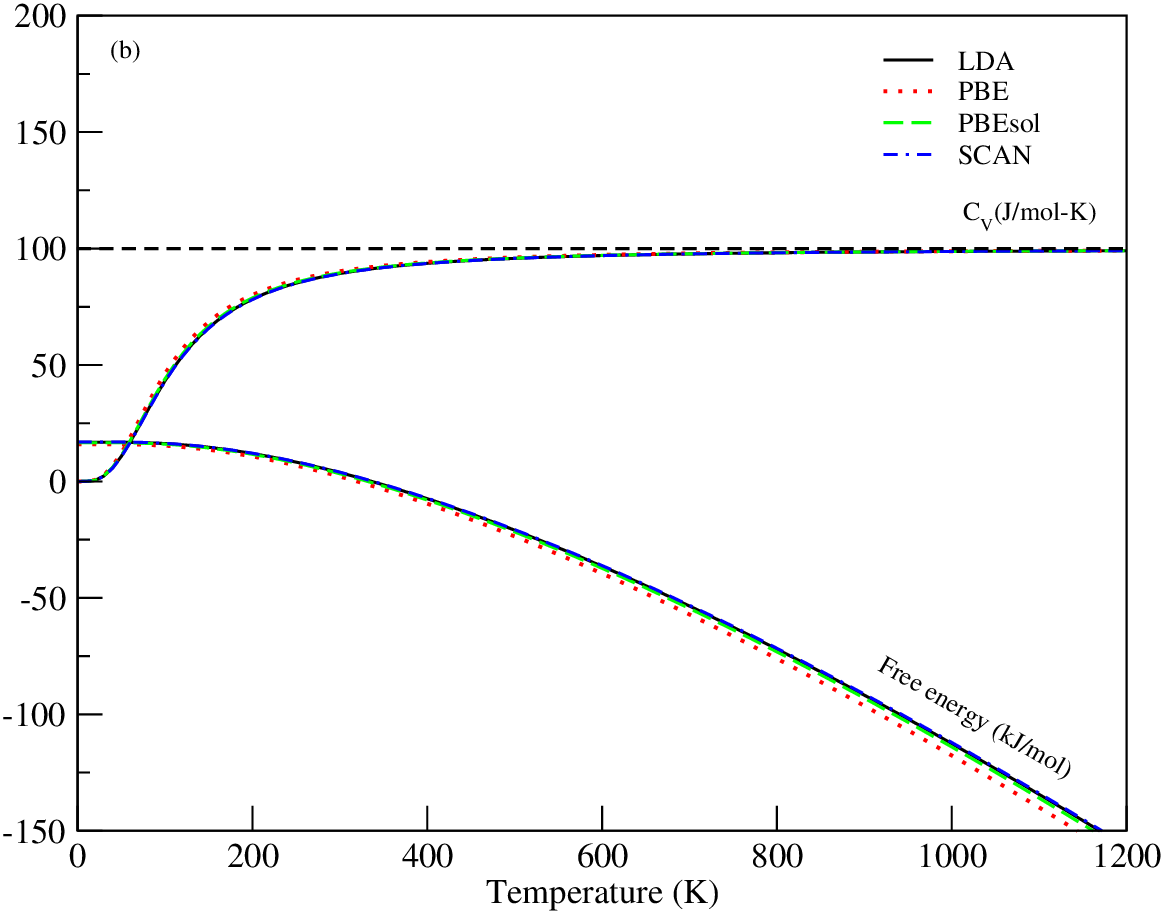}
\end{subfigure} 
\caption{Free energy (F) and constant volume specific heat ($C_{V}$) for (a) Fe$_{2}$VAl and (b) Fe$_{2}$TiSn compounds}
\label{fig:image2}
\end{figure*}

Using the thermodynamic equations, Helmholtz free energy ($F$) and constant volume specific heat ($C_{V}$) at different temperatures are calculated from PHONOPY.\cite{togo} The results obtained from all the four functionals for the two compounds are shown in Fig.4 (a) and (b), respectively. In the figure the dotted line repersents the classical Dulong and Petit value of $C_{V}$,  which is in this case $\sim$100 J/K/mol. As can be seen from the Fig. 4(a) and (b), in case of Fe$_{2}$VAl, the Dulong and Petit limit of $C_{V}$ is reaching  at $\sim$650 K. But, for Fe$_{2}$TiSn, $C_{V}$ is approaching the Dulong and Petit value at $\sim$550 K. Employing the high-temperature specific heat limiting case,\cite{ashcroft} this can be justified with the obtained higher energy (or frequency) of phonons in case of Fe$_{2}$VAl and lower energy of phonons in case of Fe$_{2}$TiSn band structures, respectively. The variation of F at different temperatures are also shown in Fig. 3 (a) and (b) for two compounds. The Kohn-Sham DFT gives the ground state energy of the static lattice.\cite{hao2012lattice} But, in a solid, the nuclei being quantum particles the vibrations exist even at zero temperature.\cite{hao2012lattice} Thus, this zero-point energy of the lattice ignored by the KS DFT is obtained using PHONOPY in the harmonic crystal approximation.\cite{hao2012lattice} The values of zero-point energies obtained in the present study for both the compounds with the help of four functionals are tabulated in Table 2. In the two figures, the intercepts to the y-axis from the Helmholtz free energy curve represent the zero-point energy.\cite{togo,ashcroft} From the plots it can be observed that the zero-point energies are $\sim$21 J and $\sim$16 J for Fe$_{2}$VAl and Fe$_{2}$TiSn, respetively. The reason for this decline in the values of zero-point energy is again related with higher and lower frequencies of the Fe$_{2}$VAl and Fe$_{2}$TiSn phonons,respectively. Because, the equational form of zero-point energy is $\frac{1}{2}\sum_{\text{\textbf{q}j}}\hbar\omega_{\textbf{q}j}$.\cite{togo,ashcroft}. Thus, the proper explanation of the variation $C_{V}$ calculated for both the compounds justify the obtained phonon energy spectrum. 
Using the thermodynamical relations, $C_{V}$ and $F$ are calculated as functions of temperature based on the equations\cite{togo}:
\begin{eqnarray}
C_V=\sum_{\bf{q}j}k_B\left(\frac{\hbar\omega_{\bf{q}j}}{k_B T}\right)^{2}\frac{exp(\hbar\omega_{\bf{q}j}/k_BT)}{[exp(\hbar\omega_{\bf{q}j}/k_BT)-1]^{2}},
\\
F=\frac{1}{2}\sum_{\bf{q}j}\hbar\omega_{\bf{q}j}+k_BT\sum_{\bf{q}j}ln[1-exp(-\hbar\omega_{\bf{q}j}/k_BT)]
\label{eq:one}.
\end{eqnarray}
Where, $k_{B}$ is Boltzmann constant, $\hbar$ is reduced Planck's constant, $\omega_{\bf{q}j}$ is phonon frequency for mode {\textbf{q},j} and $T$ is absolute temperature.

The Debye temperature $\Theta_D$ is defined as the temperature above which all modes begin to be excited, and below which they begin to frozen out.\cite{ashcroft} Looking at the phonon DOS in Fig. 2, the highest energy phonons correspond to $\sim$55 meV and $\sim$45 meV for Fe$_{2}$VAl and Fe$_{2}$TiSn, respectively based on which the $\Theta_D$ for two compounds are $\sim$660 K and $\sim$540 K. Thus, above these temperatures all the phonons are excited in the crystal lattice. In Fig. 4 (a) and (b), it can be observed that near these temperatures the $C_{V}$ is approaching Dulong and Petit limit. Which indicates that all the phonons starts to contribute to the  $C_{V}$ which justifiy the calculated phonon dispersion for two compounds.

\begin{table}[b]
\caption{\label{tab:table1}%
The zero-point energy in joule ($J$) for two compounds with four different functionals.
}
\begin{ruledtabular}
\begin{tabular}{lcc}
\textrm{Functional}&
\textrm{Fe$_{2}$VAl}&
\textrm{Fe$_{2}$TiSn}\\
\colrule
LDA & 20.99  & 16.87   \\
PBE & 20.09 & 16.02 \\
PBEsol & 20.67 & 16.63 \\
SCAN & 21.04 & 16.95 \\
\end{tabular}
\end{ruledtabular}
\end{table}

\subsection{\label{sec:level2}Density functionals and phonon properties}
In first-principles DFT calculations the total energy of a crystal is calculated as the sum of electrostatic energy, kinetic energy and exchange-correlation energy.\cite{hao2012lattice} In the direct method, the total energy of the crystal is expanded as a product of force constant matrices and displacements.\cite{yu1991force} Hence, in the harmonic theory of crystals the frequencies, in turn, energy of phonons should depend on the XC energy used to obtain the total energy. The phonons dispersion plots of Fe$_{2}$VAl obtained from LDA, PBE, PBEsol, and SCAN are shown in Fig. 1 (a)-(d), respectively. It can be noticed that, the phonon branches in PBE obtained vibrational spectrum are the mosted shifted down in energy scale by $\sim$3 meV from that of SCAN and LDA. The energies of phonon branches are nearly similar in case of LDA and SCAN meta-GGA with a difference that 6 optical branches are starting from same point at 35 meV in case of SCAN and while two degenerate branches are startig from $\sim$35 and $\sim$40 meV. While, in general, the PBEsol is producing the spectrum with energy intermediate to those from PBE and LDA (or SCAN). Observing the general features of the phonon bands for Fe$_{2}$VAl, PBE and SCAN spectra are similar with three degenerate points at $\sim$47 meV, $\sim$33 meV and 0 meV at the $\Gamma$-point. For LDA and PBEsol dispersion curves there is one more additional degenerate point $\sim$40 meV and $\sim$33 meV, repectively. These observations are well reflected in the phonon DOS of the compound in Fig. 2 (a). DOS curves from PBE are forming the lower bound but showing highest DOS in $\sim$20-40 meV region. These high values of peaks indicate that within the branches seperation is very less because of the low frequency of the phonons produced in PBE calculations. The SCAN or LDA functionals corresponding DOS curves, in general, forming the upper bound. While PBEsol phonon DOS is intermediate that of the three functionals lying in between PBE and LDA(or SCAN). 
	Fig. 1 (e)-(h) show the phonon dispersion plots for Fe$_{2}$TiSn compound from four functionals under study. In this case also, similar to the former compound, the PBE produced branches are the most shifted down in energy scale. This shift from the LDA and SCAN produced dispersions is $\sim$1 meV, which is to be noted that less than that in the former compound. The phonon branches from PBEsol functional is showing energy intermediate to those of SCAN and LDA. The general features of the four dispersion curves appears to be similar in case of this compound unlike the case of Fe$_{2}$VAl. These observations are neatly justified in the phonon DOS plot shown in Fig. 2 (b). As in the case of the former compound, PBE is forming the lower bound in the phonon energy spectrum and showing the the highest value of DOS peaks in the $\sim$10-22meV and $\sim$25-40 meV regions. The lowest peaks corresponds to the SCAN functional while LDA and PBEsol phonon DOS peaks are lying in between in the DOS peaks. But, in the enrergy, in general, SCAN and LDA are forming the upper bounds. 
 From Fig. 4 (a) and (b), it can be seen that change in XC functionals has negligible effect on $C_{V}$, while in case of F, there is small variation in higher temperatures region for the two compounds.

\section{Conclusions} 
In the present work, along with first-principles DFT to calcuate forces, phonon calculations using finite-displacement method and supercell approach are used for the study of Fe$_{2}$VAl and Fe$_{2}$tiSn compounds. The study is carried out using four XC functionals viz., LDA, PBE, PBEsol and SCAN. Under the harmonic approximation of crystals, phonon dispersion and DOS, PDOS are calculated. Using the thermodynamic relations and obtained phonon frequencies constant-volume specific heat, Helmholtz free energy as functions of temperature, zero-point energy and Debye temperatures are calculated. The obtained phonon frequencies, justify the classical Dulong and Petit limit constant-volume specific heat at higher temperatures. By comparing the phonon disperion plots and DOS from different functionals, PBE is found to give the lowest phonon energies while LDA and SCAN are giving higer values of energies. The PBE obtained dispersion energies are $\sim$2 meV and $\sim$1 meV lower than the LDA or SCAN values, respectively. The zero-point vibrational energy obtained from different functionals are in the order SCAN$>$LDA$>$PBEsol$>$PBE for both the compounds The value of $\Theta_D$ are $\sim$660 K and $\sim$540 K for the first and second compound, respectively. Slope of the acoustic branches give sound velocity which is proportional to lattice part of thermal conductivity. By examining the slopes of the acoustic branches of the two compounds, for Fe$_{2}$VAl higher value and Fe$_{2}$TiSn lower value of lattice thermal conductivity is predicted. This qualitative prediction of our work is in agreement with nature of experimental lattice thermal conductivity measurements of Leu et. al for both the compounds. Phonon spectra obtained from PBE and SCAN have similar general features while those from LDA and PBEsol have resembling features for Fe$_{2}$VAl, while this trend is not observed for the compound Fe$_{2}$TiSn.
\section{Acknowledgements}
The authors thank Science and Engineering Research Board (SERB), Department of Science and Technology, Government of India for funding this work. This work is funded under the SERB project sanction order No. EMR/2016/001511.
\section{References}

\end{document}